\documentclass[iicol,sn-basic]{sn-jnl} 
\usepackage{graphicx}
\usepackage{hyphenat}
\usepackage{amssymb,amsmath,amstext,amssymb,amsfonts}
\usepackage{epstopdf}
\usepackage{comment}
\usepackage{algorithm}
\usepackage{algorithmicx}
\usepackage{algpseudocode}
\usepackage{ifthen}
\usepackage{epsfig}
\usepackage{graphicx}
\usepackage{color,calc}
\usepackage{booktabs}
\usepackage{tabularx}
\usepackage{rotating}
\usepackage{footnote}
\usepackage{times}
\usepackage{booktabs,multirow}
\usepackage{caption}
\usepackage{subfig}
\usepackage{url}
\usepackage[multiple]{footmisc}		%Package for multiple footnotes
\usepackage[T1]{fontenc}
\usepackage{wrapfig}			%for wrapping text around figure
\usepackage{textcomp}
\usepackage{adjustbox}
\usepackage{footmisc}
\usepackage{natbib}
\jyear{2023}%
\theoremstyle{thmstyleone}%
\theoremstyle{thmstyletwo}%

\theoremstyle{thmstylethree}%

\begin{document}

\title{A Hybrid Approach For Malware Classification Using Secondary Features Fusion}


\author*[1]{\fnm{Raja Khurram} \sur{Shahzad}}\email{raja-khurram.shahzad@miun.se}
\author[2]{\fnm{Muhammad} \sur{Mustaqeem}}\email{muhammad.mustaqeem@arbisoft.com}
\author[3]{\fnm{Haroon} \sur{Elahi}}\email{haroonelahi@ieee.org}
\affil*[1]{\orgdiv{Department of Computer and System Science}, \orgname{Mid Sweden University}, \orgaddress{\city{Östersund Campus}, \state{J\"{a}mtland}, \country{Sweden}}}
\affil[2]{\orgdiv{Department of Engineering}, \orgname{Arbisoft (Pvt.) Ltd.}, \orgaddress{\city{Lahore}, \state{Punjab}, \country{Pakistan}}}
\affil[3]{\orgdiv{Department of Computing Science and Engineering}, \orgname{ Chalmers University of Technology}, \orgaddress{\city{Gothenburg}, \state{Västerbotten}, \country{Sweden}}}


\abstract{The number of malware (either variant or novel) is rapidly increasing, making malware detection and mitigation a complex problem. One approach to improving malware mitigation is automatic detection and malware family classification. However, traditional malware detection methods cannot classify detected malware into their respective families, hindering effective malware mitigation. Consequently, this paper proposes a method to automate malware detection and classification of the detected malware into respective malware families. The proposed method uses feature fusion after extracting relevant malware features such as API calls and fixed and variable length n-grams with a customized feature selection method. Moreover, for the predictive model, a voting-based approach is proposed for algorithm fusion. For the experimental evaluation of the proposed method, both binary and multi-class classification approaches are applied to the data set provided by Microsoft. Finally, the experimental results are compared with the state of the art. The experimental results indicate the effectiveness and efficiency of the proposed approach with an AUC of 0.989, accuracy of 99.72\%, and a log loss of 0.01.}

\keywords{Malware, Machine Learning, Microsoft, Feature Fusion, Algorithm Fusion}
\maketitle

\section{Introduction}\label{sec1}
Malware samples are increasing and evolving yearly due to associated monetary gains from their use in activities such as illegal content distribution and cyber attacks on organizations. For example, in the first half of 2022, 2.8 billion\footnote{https://www.sonicwall.com/medialibrary/en/white-paper/mid-year-2022-cyber-threat-report.pdf} malicious software (malware) samples were collected. Generally, most malware are variations of existing malware\footnote{https://www.av-test.org/}. To generate variations, malware authors use readily available tools that apply polymorphism, obfuscation, or a combination. These techniques change the pattern and behavior of the malware. Consequently, existing anti-malware software cannot detect variated or zero-day malware \citep{gibert_rise_of_machine_learning_2020}. Therefore, detecting and classifying malware before it serves its malicious purposes is highly prioritized. 
Effective malware mitigation requires detection, i.e., determining whether particular software is benign or malicious, and family classification, i.e., identifying the respective malware family of the detected malware. 

Malware detection is generally considered a binary problem addressed by static and dynamic analysis. The malicious file is disassembled for static analysis, and the file’s control flow is investigated for malicious patterns. On the contrary, for the dynamic analysis, the malicious file is executed in a secure environment such as a sandbox, and the file’s behavior is observed for malicious patterns. Due to the limited capabilities of static and dynamic analysis, since 2000, researchers have investigated the usage of machine learning (ML) to generalize malware detection and classification on different data sets ranging in size from small to large from different resources such as VX Heaven\footnote{Vxheaven.org} or VirusShare\footnote{https://virusshare.com}. Most data sets contain malware examples of a particular family or type. Most of these studies focus on extending static analysis by finding novel features for malware detection and increasing detection accuracy. However, these features may be used for a particular family and may not be generalized to all families. Moreover, the change in the data set may degrade the detection rate. Further, the static features may not indicate a malicious file’s behavior. In 2015, Microsoft released one of the largest disassembled malware data sets as a Kaggle\footnote{\label{kaggle}https://www.kaggle.com/c/malware-classification} challenge to improve malware mitigation. This data set can be used for a multi-class problem, where malware can be classified into different families. Researchers have used the provided data set (without any modification) to highlight ML’s importance or evaluate the proposed solution for malware classification. However, the provided data set contains only malware samples and cannot be used to classify between benign and malware.

This work modifies and extends Microsoft’s data set with benign examples and proposes an ML-based solution for malware detection (determining if a sample is benign or malicious) and, subsequently, the detected malware’s classification to respective families. For the experimental purpose, the Microsoft data set is partially used to evaluate the proposed approach for malware classification. The proposed approach addresses two distinct stages of ML, i.e., feature engineering and modeling.

The contributions of this article are as follows:
\begin{itemize}
\item This study extends the Microsoft data set with benign examples to address binary and multi-class problems.
\item This study uses application programming interface (API) calls, Dynamic Link Library (DLL) imports, and operation code mnemonics (OpCodes) as primary features.
The combinational analysis of API calls and DLL imports is also performed to obtain high accuracy and valuable features. OpCodes are used to generate secondary features, i.e., fixed length (\textit{n}) and \textit{variable length}-grams. All these features are combined to generate a feature set. 
\item Different feature selection methods, such as filter, wrapper, and hybrid methods, are evaluated. Further, for feature selection, a customized backward selection algorithm is proposed.

\item After the feature selection, feature fusion is performed to get a data set with the best features. As per the authors’ best knowledge, no feature fusion representing all families of feature selection methods is investigated for malware classification.

\item Algorithm Fusion is performed to suggest a weighted voting-based ensemble to determine the outcome.

%\item  
\end{itemize}

%===========================================================
%==========================================================
\begin{figure*}
  \includegraphics[width=\textwidth,height=6cm]{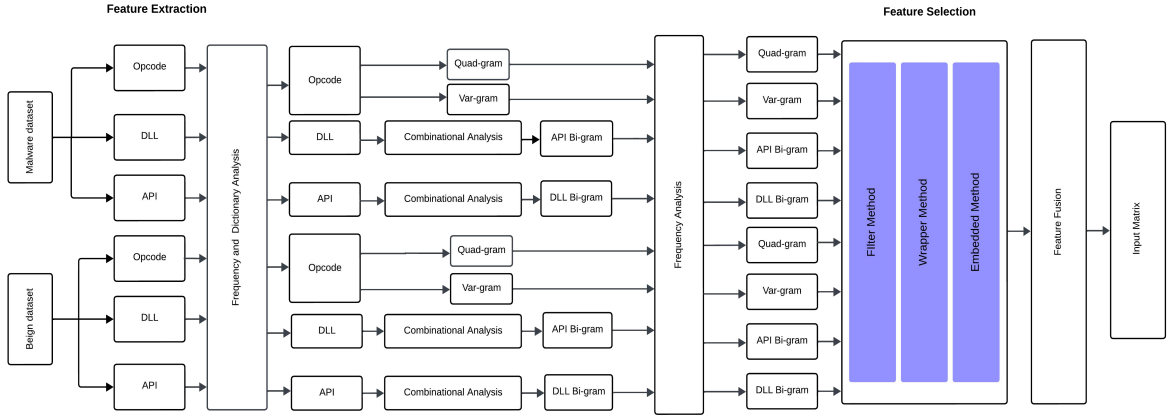}
  \caption{Feature Selection Process}
  \label{figMicrosoft:FeatureSelection}
\end{figure*}
%===========================================================
%===========================================================
%===========================================================
%===========================================================
\subsection{Related Work}
%===========================================================
%===========================================================

For malware detection, the cost of misclassifying a malware is higher than the cost of misclassification of a benign file due to its impacts on the system. Thus, researchers have used machine learning methods such as feature engineering, feature selection, and generating ensembles to improve malware detection and classification accuracy on different platforms. For the Microsoft malware data set, two types of studies are conducted, i.e., highlighting the importance of machine learning for malware classification and evaluating the proposed approach \citep{ronen2018microsoft}. The latter group may be further divided into two categories: evaluation of the proposed feature engineering approach and assessment or comparison of predictive models. Different studies have presented their work for extracting novel features and reducing features dimensionality \citep{TANG_2023,WANG_2022,Euh_20,Chen_20,Hu_2019}.  
However, a few authors have reported that using both modalities, i.e., byte sequence and disassembled code, may cause overfitting \citep{Gibert_Orthurs_20}. For the feature selection and dimensionality reduction step, different authors have used different methods \citep{GUO_2023,ZHU_2022,LI_2022,GIBERT_2022,ahmadi2016novel, Unal_20, Mark_21}.    
For the classification, algorithms from different classes are used, i.e., AdaBoost, XGBoost, LG Boost, Random forest, Extra trees, Rotation trees, and Random Forest \citep{Euh_20}. A Few authors have reported customized ensembles and parameter tuning \citep{Chen_20, Gibert_Orthurs_20}. 
In an attempt to select the best features, the researchers have used eight different filter-based feature selection methods \citep{Sahin2021}. Four filter methods commonly used in studies are information Gain, Odds Ratio, Chi-Square, and Inverse document frequency. The new methods used by reaserahers are Document frequency thresholding, M2, and Relevance frequency feature selection. Later, for the modeling, they used K-Nearest Neighbors (\textit{k}NN), Naive Bayes (NB), Sequential Minimal Optimization (SMO), MLP, Random Forest (RF), C4.5, and Logistic Regression (LR) algorithms. The highest performance is 0.955, which is achieved with the SMO algorithm.

The Microsoft dataset is not only used for malware classification or as a stand-alone data set \cite{NAEEM_2023,ALANI_2023, RUSTAM_2023,JOYCE_2023,MIMURA_2023}. In an attempt to reduce features, the researchers have used the data set generated by combining heartbeat and threat reports collected by Microsoft’s endpoint protection solution \citep{Pan_20}. 
In another study, the authors have presented their work to classify a new malware family based on small historical samples using text classification methods \citep{Ding_Self_20}. The authors used the Microsoft data set, which inherits the properties of an imbalanced data set, 
to simulate two different scenarios. Their proposed method shows reasonable accuracy in detecting the new variant. However, their method is time-consuming.

%%%%%%%%%%%%%%%%%%%%%%%%%%%%%%%%%%%%%%%%%%

%===========================================================
%===========================================================
\section{Experimental Setup}
%===========================================================
%===========================================================
\subsection{Data set}
Microsoft released the experimental data set for a challenge at Kaggle, titled "\textit{Microsoft Malware Classification Challenge}"\footref{kaggle}\footnote{https://www.kaggle.com/c/malware-classification}. The  data set contains nine different malware families: Ramnit, Lollipop, Kelihos\_ver3, Vundo, Simda, Tracur, Kelihos\_ver1, Obfuscator.ACY, and Gatak.  

The data set contains both training and test data. The training data set consists of 186 GB of data. The training data set contains 10,868 files. The test data size is 189 GB, and contains 10873 examples. The released data set contains two types of disassembled files for each malware sample, i.e., byte code, which contains the hexadecimal dump (47.3 GB of train and test data respectively ) of malware files and .asm files, which are disassembled files in assembly instructions form. The .asm files are obtained by using IDA Pro\footnote{https://www.hex-rays.com/products/ida/}. 
In this particular study, for the experimental purpose, the data set represented as .asm files is used because it is unsure how the hexadecimal files are generated. Further, the data set is modified by removing the obfuscated files and extended by adding 1,609 benign disassembled files (.asm). The benign files are downloaded from download.com\footnote{https://download.cnet.com/}. The downloaded files are disassembled by using IDA Pro to obtain .asm files. The purpose of including benign files is multi-fold. The first purpose is to change the multi-class classification problem to a binary classification problem between malware (all families) and benign, and the other fold is to recognize the exact family of the malware and distinct it from benign files. The modified data set inherited the class imbalance problem \citep{gibert_rise_of_machine_learning_2020}. For example, the Ramnit class has only 1541 instances out of 10868. To avoid class imbalance, stratified random sampling with replacement is used. It is worth noting that the data set is available on request from the authors. 
%===========================================================
%===========================================================
\subsection{Feature Extraction}
A .asm file is divided into different sections such as .data section, .text section, .idata section, etc. \citep{Malware_Analysis_Book_2012, gibert_rise_of_machine_learning_2020} Each section has particular features. However, the .text section is of particular interest. The .text section contains different types of valuable features/information, such as assembly instructions, hexadecimal code of each byte with memory segment, readable strings, API calls, DLL imports, headers information, etc.
Previous studies are not conclusive about the best features. However, it is suggested that API calls, DLL imports, and assembly instructions are better features as an input for learning algorithms for the classification \citep{Liu_2017, Chen_2017, shabtai_2009} compared to hexadecimal code and readable strings. Thus, three features, i.e., API Calls, DLL imports, and assembly instructions, are extracted to use as a feature by writing a parser in Python\footnote{www.python.org} (and regex). 

The extracted assembly instructions consist of two parts, i.e., OpCode and operands. The operands are architecture-dependent and cannot provide helpful information. One of the studies has suggested that the registers' frequency of use may be used as a feature for malware detection \citep{ahmadi2016novel}. However, the generalizability of the suggested approach is unsure. Thus, operands from the assembly instructions are discarded, and OpCodes are extracted in the form of \textit{uni}-gram. The extracted OpCodes are saved in the order of occurrences. It is worth noting that the extracted OpCodes, API Calls, and DLL imports are saved as separate feature sets in order of their occurrence. 
%===========================================================
%===========================================================
\subsection{Features}
%===========================================================
%===========================================================
\label{subsec:features}

\textit{N}-gram is a contiguous sequence of \textit{n} items (e.g., words) (extracted from a data set), where \textit{n} may range from one to \textit{n} \citep{Witten2016}. For experimental purposes, the extracted feature sets contain the OpCodes, API calls, and DLL\footnote{https://docs.microsoft.com/en-us/troubleshoot/windows-client/deployment/dynamic-link-library} imports in textual form, and each feature can be considered a word. 
%=======================
\paragraph{OpCode:} Each extracted OpCode may be considered a \textit{uni}-gram, which may not provide helpful information about the file's structure and functionality (purpose) \citep{shahzad_adware_2011}. Moreover, the extracted OpCode data set contains noise, i.e., unofficial or customized OpCodes. Thus, to remove the noise from extracted OpCodes, dictionary-based primary feature selection is applied (Please,  refer to section \ref{subsec:primaryfeatureselection}). Further, to represent features in the input matrix, the \textit{uni}-gram OpCodes are used to generate \textit{quad}-grams and \textit{variable-length} grams. In previous studies,  \textit{quad}-grams and \textit{variable-length} grams have shown promising results compared to other sizes \citep{shahzad_adware_2011}. However, to the best of the authors' knowledge, both \textit{quad}-gram and \textit{variable-length} n-grams have not been used together nor have their performances been compared.

\paragraph{API and DLL}
Application programming interface or Windows API provides an integration interface for users or software to communicate and integrate. A DLL file contains a program or a collection of programs, generally referred to as a library with relevant data or a data structure that other programs can use. For the malware classification task, it is suggested that API and DLL names are crucial features of the software, which may provide valuable information. However, for generating API and DLL \textit{n}-grams, previous studies are not conclusive for optimal size \citep{Hu_2019, Attaluri2009} and have used them as a \textit{uni}-gram. Thus, to find the optimal \textit{n}-gram size, the combinational analysis of the API and DLL data sets is performed. The results have suggested that a \textit{bi}-gram is an optimal size. There is an ignorable difference in the classification accuracy if \textit{tri}-gram, \textit{quad}-gram or larger \textit{n}-grams are used as input compared to \textit{bi}-gram. However, the computational cost is relatively high with each increasing size. Thus, the API and DLL data sets are used to generate API and DLL \textit{bi}-grams as features for the input matrix.

%\begin{algorithm}
%	\caption{Proposed Backward Selection Algorithms}
%	\begin{algorithmic} 
%		%\Function{BackwardSelection}{$TrainingSet$}
%		\Procedure{BackwardSelection} {$TrainSet,TargetClass,MinFeatures$}
%		\Require $MaximumAccuracy \gets 0, BestFeatures \gets \emptyset, Train\gets \emptyset, Test\gets \emptyset, Accuracy \gets 0$ %\{\} or $\O$
%		\Repeat 
%		\State 	$Train \gets Split(TrainSet)$  %\Comment{Training set is split into 80/20 by %using sampling with replacement}
%		\State 	$Test \gets Split(TrainSet)$
%		\State  $Accuracy \gets CalculateAccuracy$
%		\If{$Accuracy > MaximumAccuracy$} 
%		\State $MaximumAccuracy \gets Accuracy$
%		\State $Update(BestFeatures)$
%		\EndIf
%		\State $FeaturesImportance \gets CalculateFeaturesImportance(Train)$
%		\State $TrainSet \gets RemoveLeastImportantFeatures(TrainSet)$
%		\Until{$NumberOfRemainingFeatures \geq MinFeatures$}
%		\State \Return $BestFeatures$
%		\EndProcedure 
%		\label{BackwardAlgo}
%		%\EndFunction 
%	\end{algorithmic}
%\end{algorithm}

\begin{figure}[!t] %for floating picture use *
	\centering
	\includegraphics[width=\linewidth,height=6cm]{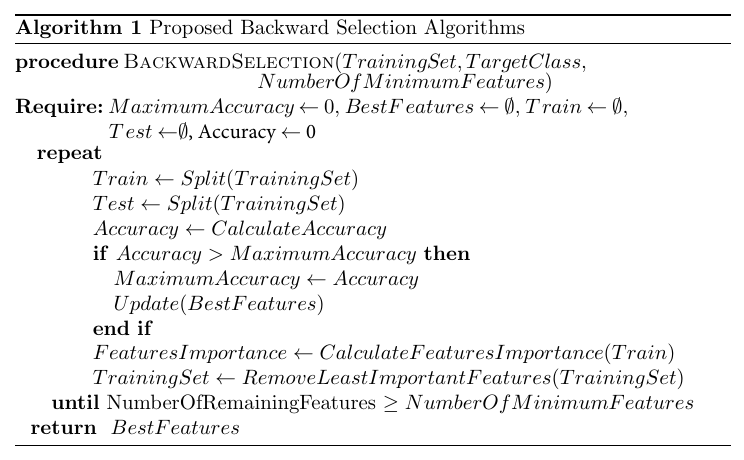}
	\label{BackwardAlgo}
\end{figure}
\subsection{Feature Selection}
Feature selection is generally considered a one-step problem. However, for this particular study, after a manual inspection of the data, feature selection is performed in two stages, i.e., primary and secondary feature selection. The primary feature selection is further divided into two stages, i.e., dictionary-based selection and frequency analysis. After the primary feature selection, each file's representative features (Please,  refer to section \ref{subsec:features}) are generated and used as input for the secondary feature selection. For the secondary feature selection, statistical measures and different machine learning based methods are used. The purpose of dividing one-step feature selection into two steps is to obtain the most valuable features. Additionally, each learning algorithm has its own inductive bias, meaning that the best features of each algorithm are different. Combining the best features obtained from different methods may help create a comprehensive input matrix and improve classifier performance. The feature selection process is explained in Figure \ref{figMicrosoft:FeatureSelection}.  
\subsubsection{Primary Feature Selection}
\label{subsec:primaryfeatureselection}
The primary feature selection is performed to remove irregular and rarely used features from the extracted feature set and reduce the data set size significantly. 
\paragraph{Dictionary-based Selection}
\label{subsec:dictionary}
The 8086 architecture\footnote{https://www.intel.com/}
manuals by Intel and Microsoft websites are consulted to develop the dictionaries of regular OpCodes, API calls\footnote{https://docs.microsoft.com/}. 
Intel's architecture guideline contains 1094 regular OpCodes. After comparing OpCodes with the dictionary, many irregular OpCodes are found and discarded. A majority of irregular OpCodes are custom made OpCodes. After removing the irregular OpCodes, the reduced OpCode set contains standard 539 OpCodes per Intel's guidelines. For the API calls and DLL imports, the total number of standard APIs is 8,068, and DLLs are 1,957. However, no non-standard API call or DLL import is found in the data set. 

\paragraph{Frequency Analysis}
Frequency analysis is performed at two distinct stages. First, after the feature extraction, i.e., extracting OpCodes, APIs and DLLs from disassembled files. Secondly, after generating feature representation for the input matrix such as \textit{quad}-gram. For the first stage, a frequency analysis of extracted features is performed to find the most commonly used features and remove rarely used features. For this purpose, a hash table is generated for every feature (OpCode, API, and DLL) for mapping their frequencies. For OpCodes, there is a significant decrease in frequency after the first 300 \textit{uni}-grams.  For the API features, the frequency analysis has suggested that after $1329^{th}$ API, the frequency of API calls is 100.  After $3,320^{th}$ API, the frequency of API calls is ten or less than ten, and after $6,341^{th}$ API, the frequency of API calls is only one. For DLLs, after the first 86 DLLs, the frequency count is 100, and after $434^{th}$ DLL, the frequency of DLL import is ten or less than ten. 

After generating feature representations, a secondary frequency analysis is performed to decide a threshold for generated representations, which suggests a frequency threshold of 50. Thus, all features with a frequency of less than 50 are discarded.

%===========================================================
%===========================================================
\subsubsection{Secondary Feature Selection}
For experimental purposes, this article focuses on three supervised feature selection strategies, i.e., filter, wrapper, and hybrid \citep{Noelia2007}.  

\paragraph{Filter method}
Filter methods use uni-variate statistics to identify/measure the relevance of features by their correlation with the target variable without induction of a learning algorithm \citep{kuhn2013applied, Noelia2007}. 
To represent filter methods, Shannon Entropy is a widely used method for the feature selection \citep{Witten2016}. Entropy is a measure of randomness in the feature's possible outcome. Entropy is computed on all feature representations of each malware and benign sample.  
%===========================================================

%===========================================================
\paragraph{Wrapper Method} 
Wrapper methods convert the feature selection problem into a search problem using a learning algorithm. Wrapper methods can have either a forward feature selection process or a backward feature selection/elimination \citep{Noelia_2009,chowdhury_variable_2020}.  For this particular study, we propose a backward feature selection algorithm, as presented in Algorithm 1 and further explained below. 

In the previous studies, the Random Forest (RF) algorithm has shown promising results for feature selection and malware classification \citep{Hu_2019}. Another algorithm, Regularized Greedy Forest (RGF) has shown promising results in different Kaggle's competition \citep{Rie2014}. Thus, we used both Random Forest and Regularized Greedy Forest algorithm\footnote{https://github.com/RGF-team/rgf} as base algorithms for backward feature elimination. Both algorithms have used 10-fold cross-validation. For the hyperparameter tuning to improve ensemble performance, giving the manual value to both algorithms can be costly in terms of time and resources. Thus, an automated process is used to give a predefined list of parameter values, and the best parameters are selected and used for the feature selection. Finally, for the RF, 100 trees are generated without a depth limit. It is suggested that an RF with 100 trees may provide maximum accuracy \citep{Shahzad_consensus_2015}. If the number of trees is grown beyond this limit, they may increase the computation cost without significantly increasing accuracy. The Gini impurity is used for splitting the nodes, which is a criterion for calculating the information gain of a feature, and its value lies between 0 to 0.5 \citep{Witten2016}. For RGF, a variation of RGF, i.e.,  RGF Sib algorithm\footnote{https://www.kaggle.com/carlmcbrideellis/introduction-to-the-regularized-greedy-forest}, which uses "\textit{minimum penalty regularization with the sum-to-zero sibling constraints}" is used as a base algorithm. For the loss calculation, the square loss (LS) method is used, which can be calculated as follows: 
\begin{equation}%\nonumber 
	Square Loss = (p-y)^2/2 
\end{equation}
Moreover, a limitation of a maximum of 1000 leaf nodes is applied to the algorithm. 

%===========================================================
\paragraph{Embedded Method}
Embedded methods combine the features of filters and wrapper methods. 
To represent the embedded methods, Lasso and eXtreme Gradient Boosting (XGBoost)\footnote{https://github.com/dmlc/xgboost/} algorithm is used. XGBoost is also a decision tree-based ensemble, which is designed for speed, flexibility and performance \citep{Chen_2016}.

%===================================================================================
\begin{figure}[htbp] %for floating picture use *
	\centering
	\includegraphics[width=\linewidth,height=5cm]{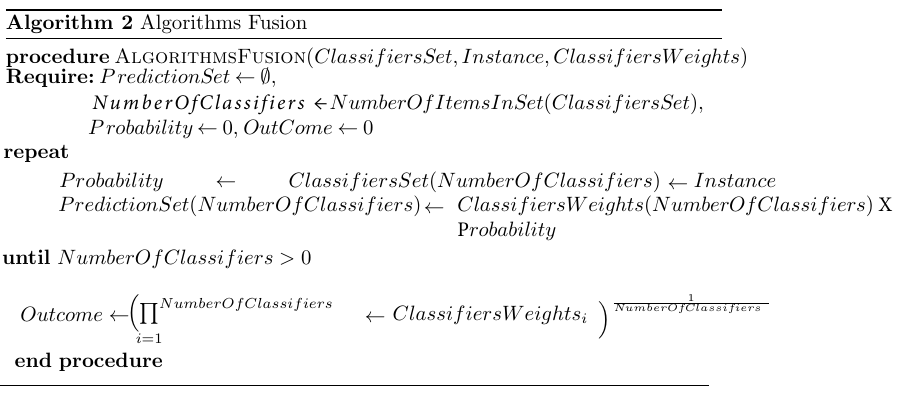}
	%\caption{Confusion Matrix}
	\label{AlgoFusion}
\end{figure}
%===================================================================================
%===========================================================
%===========================================================

%\vspace{-1cm}
\subsection{Feature Fusion}
%===========================================================
%===========================================================
\label{sec:FeatureFusion}
Feature fusion integrates multiple feature sets to obtain a single feature set, which can improve accuracy. Different feature fusion methods are suggested to stack all features in a single feature vector. Thus, to combine the best features from all representations, a union of all features subset is taken.  
\begin{table*}[!htbp]
	\centering
	\caption{Multi-class Experiment's Results}
	\label{tab:MultiClass}
	\resizebox{.5\textwidth}{!}{% <------ Don't forget this %
		\begin{tabular}{|l|l|c|c|c|}
			\hline
			\textbf{Classifiers}                           & \textbf{Features}                    & \textbf{Selected Features} & \textbf{Accuracy} (\%) &\textbf{ T. Time}* \\
			\hline
                \multirow{4}{*}{CART}  & DLL \textit{bi}-gram                 & 1306 & 95.76 & 0.015  \\
			& API \textit{bi}-gram                 & 5371 & 96.28 & 0.5    \\
			&  \textit{variable-length} Gram & 2659 & 97.24 & 0.25   \\
			&  \textit{quad}-gram            & 3466 & 94.76 & 0.3   \\
			\hline
   			\multirow{4}{*}{Logistic Regression}   & DLL \textit{bi}-gram                  & 1306  & 99.1                  & 1.4                     \\
			& API \textit{bi}-gram                  & 5371  & 96.98                 & 2.33                    \\
			&  \textit{variable-length} Gram  & 2659  & 99.45                 & 3.5                     \\
			&  \textit{quad}-gram             & 3466  & 99.4                  & 2.7                     \\
			\hline
   			\multirow{4}{*}{SVM}                                   & DLL \textit{bi}-gram                  & 1306  & 98.11                 & 89.4                    \\
			& API \textit{bi}-gram                  & 5371  & 95.32                 & 421.38                  \\
			&  \textit{variable-length} Gram  & 2659  & 98.19                 & 300.5                   \\
			&  \textit{quad}-gram             & 3466  & 97.2                  & 248.13                  \\
			\hline
   			\multirow{4}{*}{\textit{k}NN}                    & DLL \textit{bi}-gram                  & 1306  & 96.4                  & 18.55                 \\
			& API \textit{bi}-gram                  & 5371  & 92.33                 & 60.87                   \\
			&  \textit{variable-length} Gram  & 2659  & 96.8                  & 44.49                   \\
			&  \textit{quad}-gram             & 3466  & 97.78                 & 50.4                    \\
			\hline
   			\multirow{4}{*}{Naive Bayes}         & DLL \textit{bi}-gram                 & 1306 & 94.14 & 25.25  \\
			& API \textit{bi}-gram                 & 5371 & 91.2  & 123.15 \\
			&  \textit{variable-length} Gram & 2659 & 95.53 & 43.14  \\
			&  \textit{quad}-gram            & 3466 & 97.7  & 52.16  \\
			\hline     
   			\multirow{4}{*}{Neural Network}        & DLL \textit{bi}-gram                  & 1306  & 98.2                  & 3.15                    \\
			& API \textit{bi}-gram                  & 5371  & 96.12                 & 10                      \\
			&  \textit{variable-length} Gram  & 2659  & 99.68                 & 6.5                     \\
			&  \textit{quad}-gram             & 3466  & 99                    & 7.25                    \\
			\hline
   			\multirow{4}{*}{Random Forest}       & DLL \textit{bi}-gram                  & 1306  & 97.3                  & 0.01                    \\
			& API \textit{bi}-gram                  & 5371  & 95.43                 & 1                       \\
			&  \textit{variable-length} Gram & 2659  & 99.72                 & 0.3                     \\
			&  \textit{quad}-gram             & 3466  & 98.9                  & 0.1                     \\
			\hline
            \multirow{4}{*}{Adaboost}  & DLL \textit{bi}-gram                  & 1306  & 43.43                 & 8.44                    \\
			& API \textit{bi}-gram                  & 5371  & 34.56                 & 24.1                    \\
			&  \textit{variable-length} Gram & 2659  & 95.45                 & 10.98                   \\
			&  \textit{quad}-gram             & 3466  & 90.2                  & 18                      \\
			\hline
   			\multirow{4}{*}{XGBoost}   & DLL \textit{bi}-gram                  & 1306  & 97.3                  & 15.25                   \\
			& API \textit{bi}-gram                  & 5371  & 95.43                 & 40.56                   \\
			& \textit{variable-length} Gram & 2659  & 99.5                  & 21.34                   \\
			&  \textit{quad}-gram             & 3466  & 98.9                  & 30.28                   \\
			\hline
			\multirow{4}{*}{Light Gradient Boosting}               & DLL \textit{bi}-gram                  & 1306  & 99.1                  & 10                      \\
			& API \textit{bi}-gram                  & 5371  & 98.4                  & 22.4                    \\
			&  \textit{variable-length} Gram  & 2659  & 99.6                  & 16.34                   \\
			&  \textit{quad}-gram             & 3466  & 99.4                  & 19.19                   \\
			\hline
   \multicolumn{5}{l}{*The given training time is in minutes.}
	\end{tabular} }

\end{table*}

%===========================================================
%===========================================================
\subsection{Algorithms}
%===========================================================
%===========================================================
Two types of supervised algorithms are used for experimental purposes, i.e., base/basic algorithms and tree-based ensembles 
\citep{Sagi_Ensemble_learning, Dong_ensemble_2020}. The basic algorithms used in the experiment are Classification and Regression Trees (CART), which is a decision tree-based algorithm, Naive Bayes (NB) algorithm, which is based on Baye’s theorem and is used to generate a probabilistic classifier, Support Vector Machine (SVM) which is a robust learning algorithm based on Vapnik–Chervonenkis theory, Logistic Regression (LR) which is based on the concept of probability and uses a logistic function, \textit{k}-Nearest Neighbors (\textit{k}NN), which assumes that similar observations are present in close proximity and a basic neural network algorithm %\citep{Witten2016,Fix_KNN_1989,Altman_KNN_1992}
\citep{Witten2016}. For the tree-based ensembles, Random Forests (RF), which is an ensemble of decision trees, decides about the class using majority voting; %Bagging, which is a meta-learning ensemble algorithm and uses bootstrap aggregating (Bagging), and 
Boosting, which is a meta-learning ensemble and combines weak learners,  are used %\citep{Witten2016,breiman2001random} 
\citep{Witten2016}. CART is used as a base algorithm for RF (with 100 trees) and Bagging. Boosting can be either adaptive Boosting (AdaBoost) or Gradient Boosting (GBoost). Two variants of GBoost are eXtreme Gradient Boosting and Light Gradient Boosting Machine (LightGBM). The LightGBM differs from other GBoost algorithms in growing the tree criteria. Moreover, the configuration used in algorithms is as follows: The XGBoost and LGBoost are used with an estimator of 100, Logistic Regression is used with an alpha of 0.0001 and a limit of 10000 iterations; Neural Network is used with a single layer and a limit of 300 iterations;  Random Forest is used with an estimator of 200. The pruning factor for CART is 0.001 and the number of neighbors for \textit{k}NN is five. It is worth noting that the majority of these algorithms have used parallel processing on all cores. 

%===========================================================

%\begin{table}
%	\setlength{\extrarowheight}{2pt}
%	\begin{NiceTabular}{c*{6}{r}}[hvlines]
%		\Block{2-1}{Dataset} &
%		\Block{1-2}{A} & &
%		\Block{1-2}{B} & & 
%		\Block{1-2}{C} \\
%		& {O.B.R} & \Block[c]{1-1}{A.R} & {O.B.R} & \Block[c]{1-1}{A.R} & {O.B.R} & \Block[c]{1-1}{A.R} \\
%		D1 & 2.1 \%& 2.1 \%& 2.1 \%& 2.1 \%& 2.1 \%& 2.1 \%\\
%		D2 & 11.6 \%& 11.6 \%& 11.6 \%& 11.6 \%& 11.6 \%& 11.6 \%\\
%		D3 & 5.5 \%& 5.5 \%& 5.5 \%& 5.5 \%& 5.5 \%& 5.5 \%\\
%	\end{NiceTabular}
%\end{table}

\begin{table*}[ht]
	\centering
	\caption{Results of Top 5 Classifiers}
	\label{tab:allResults}
	\resizebox{\textwidth}{!}{% <------ Don't forget this %
		\begin{tabular}{|l|l|l|l|l|l|l|l|l|l|}
			\hline
			
		%	\multicolumn{4}{|c|}{} & \multicolumn{2}{c|}{Precision} &       \multicolumn{2}{c|}{Recall} &       \multicolumn{2}{c|}{  F1  }    \\ \hline

		%	\multirow{2}{*}{Classifiers}             & 	\multirow{2}{*}{Log Loss} & 	\multirow{2}{*}{Accuracy} & 	\multirow{2}{*}{AUC}   & \multicolumn{2}{c|}{Precision} &       \multicolumn{2}{c|}{Recall} &       \multicolumn{2}{c|}{F1} \\ 
			%\cline{5-10}

		\multirow{2}{*}{\textbf{Classifiers}}             & 	\multirow{2}{*}{\textbf{Log Loss}} & 	\multirow{2}{*}{\textbf{Accuracy}} & 	\multirow{2}{*}{\textbf{AUC}}   & \multicolumn{2}{c|}{\underline{\textbf{Precision}}} &       \multicolumn{2}{c|}{\underline{\textbf{Recall}}} &       \multicolumn{2}{c|}{\underline{  \textbf{F1}  }}    \\ 
	%\cline{5-6}
	
		%	Classifiers             & Log Loss & Accuracy & AUC   & Precision &       & Recall &       & F1    &       \\
		%Classifiers	&       Log Loss    &    Accuracy     &  AUC     & Micro     & Macro & Micro  & Macro & Micro & Macro \\ \hline
			&            &         &       & Micro     & Macro & Micro  & Macro & Micro & Macro \\ \hline
		
		%	Ensemble (Proposed)      & 0.052    & 98.874 (0.000005)  & 0.989 & 0.989     & 0.93  & 0.989  & 0.979 & 0.989 & 0.936 \\ 	\hline
			XGBoost                 & 0.048    & 98.773 (0.05694)  & 0.987 & 0.988     & 0.924 & 0.988  & 0.975 & 0.988 & 0.932 \\ 	\hline
			Random Forest           & 0.075    & 98.59 (0.05475)    & 0.986 & 0.986     & 0.925 & 0.986  & 0.972 & 0.986 & 0.928 \\ 	\hline
			Light Gradient Boosting & 0.075    & 98.736 (0.02856)  & 0.986 & 0.988     & 0.927 & 0.988  & 0.973 & 0.988 & 0.931 \\ 	\hline
		Logistic Regression     & 0.043    & 98.819 (0.05887)  & 0.992 & 0.989     & 0.927 & 0.989  & 0.985 & 0.989 & 0.938 \\ 	\hline
			Neural Network          & 0.075    & 98.736 (0.02856)  & 0.988 & 0.988     & 0.927 & 0.988  & 0.973 & 0.988 & 0.931 \\
			\hline
	\end{tabular} }
\end{table*}

%===========================================================
%===========================================================
\subsection{Algorithms Fusion}
\label{sec:AlgorithmFusion}
%===========================================================
%===========================================================
Generally, an ensemble is created by combining a finite set of learning algorithms. An ensemble may contain either homogeneous or heterogeneous algorithms, which may be trained on a subset of data, and their predictions are used to determine the outcome of ensemble \citep{Witten2016}. In our experiment, heterogeneous algorithms are used to generate an ensemble. Some of them are learning algorithms with their own biases, and some of them are ensemble in their nature. Thus, an algorithm fusion is proposed to combine decisions from various algorithms and determine the outcome, as shown in Algorithm 2. The suggested algorithm fusion is performed in different stages as follows:

\begin{itemize}
	\item The weight for each classifier is calculated. For calculating the weight, the probability matrix of the test data set is used. The Sequential Least Squares Programming (SLSQP) algorithm, which is a sequential quadratic programming (SQP) algorithm\footnote{http://degenerateconic.com/slsqp.html}, is used to determine the weight within a range. The value of 0.5 is given as a minimal value. The probabilities table is multiplied by the selected weight. The log loss (Please, see the equation \ref{eqMicrosoft:logloss}) of the resultant table is determined. This process is repeated by changing the weight suggested by the SLSQP method until the log loss no longer decreases. The last weight that affected the log loss is considered the final weight for the classifier. This process is repeated for each classifier.   
	\begin{equation}
		Log Loss = - \frac{1}{N}\sum_{i}^{M}\sum_{j}^{M} y_{ij}log(p_{ij}
            \label{eqMicrosoft:logloss}
	\end{equation}
	\item When an instance is given to the ensemble for determining its class. Each classifier provides its probabilities for each class. Further, a two-dimensional array is generated, containing all classifiers in rows, and against each classifier, its probabilities for each class are saved in columns. 
	\item The rows are multiplied with each classifier's weight, generating a new result matrix. 
	\item The geometric mean of each column is calculated, resulting in a single array with ten probabilities. 
	\item The maximum geometric mean is used as the outcome for the given instance.  
\end{itemize}

%===================================================================================

%%\begin{algorithm}
%	\caption{Algorithms Fusion}
%	\begin{algorithmic} 
%		%\Function{BackwardSelection}{$TrainingSet$}
%		\Procedure{AlgorithmsFusion}{$ClassifiersSet, Instance, ClassifiersWeights$}
%		\Require $PredictionSet \gets \emptyset,  
%		NumberOfClassifiers \gets NumberOfItemsInSet(ClassifiersSet), Probability \gets 0, OutCome \gets 0$
%		\Repeat
%		
  %      \State $Probability \gets ClassifiersSet(NumberOfClassifiers) \gets Instance$
%		\State $PredictionSet(NumberOfClassifiers) \gets ClassifiersWeights(NumberOfClassifiers) \times Probability$
%		\Until{NumberOfClassifiers > 0}
%		\state $Outcome \gets \left(\prod _{i=1}^{NumberOfClassifiers}ClassifiersWeights_{i}\right)^{\frac {1}{NumberOfClassifiers}}$ %\algorithmiccomment{Comment}
%		
%		
%		\EndProcedure 
%		\label{AlgoFusion}
%		%\EndFunction 
%	\end{algorithmic}
%\end{algorithm}
%===========================================================
%===========================================================
%=====================================================================
\begin{figure}[!t] %for floating picture use *
	\centering
	\includegraphics[width=\linewidth,height=8cm]{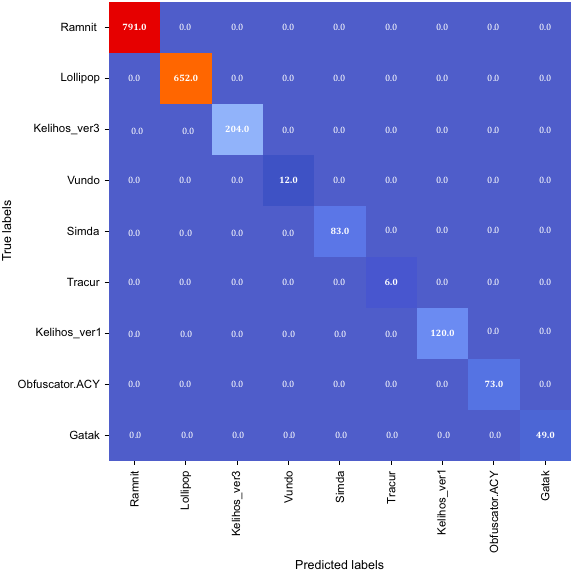}
	\caption{Confusion Matrix}
	\label{figSpy:ConfusionMatrixOfEnsemble}
\end{figure}
%=====================================================================

\subsection{Evaluation Measures}\label{subsec:Eval}
%===========================================================
%===========================================================
The performance of each learning algorithm is evaluated by performing 10-fold Cross-Validation. 
Moreover, confusion matrices are generated by using the responses from classifiers. The four parameters to generate a confusion matrix are as follows: True Positive (TP) represents the correctly identified malware programs and their families. False Positive (FP) represents the incorrectly classified benign programs. True Negative (TN) represents the correctly identified benign programs, and False Negative (FN) represents the incorrectly identified malware programs and their families. The performance of each classifier is evaluated using Detection Rate (DR), which is the ratio of malware programs correctly identified from the total number of malware programs; False Alarm Rate (FAR), the ratio of malware programs incorrectly identified; Precision, which is a ratio of correct positive predictions to the total number of positive predictions and measure of quality (Please,  see the equation \ref{eq:Pre}), Recall, which is a measure of quantity and is ratio of identified positive over all the positives in the data set (Please,  see the equation \ref{eq:Rec}). For aggregated indicators, Accuracy (ACC), the percentage of correctly identified programs (Please,  see the equation \ref{eq:Acc}); and  F-score (F1), which is a harmonic mean of Precision and Recall (Please,  see the equation \ref{eq:F1}), are used. The last evaluation measure is the Area Under the Receiver Operating Characteristic Curve (AUC), a single-point value derived from an ROC curve. The higher AUC of an algorithm indicates that the algorithm is more robust and better in classification. Another evaluation measure is log loss \citep{Vovk2015}. Kaggle recommends log loss for comparing models' performance. The log loss measures the uncertainty of the predicted probabilities by comparing them with the corrected probabilities of a given model. Thus, a lower log loss indicates a better performance by the model.

\begin{equation}
	Precision = \frac{TP}{TP+FP}
 \label{eq:Pre}
\end{equation}

\begin{equation}
	Recall = \frac{TP}{TP+FN}
 \label{eq:Rec}
\end{equation}
\begin{equation}
    Accuracy = \frac{TP+TN}{TP+TN+FP+FN}
    \label{eq:Acc}
\end{equation}

\begin{equation}
	F1 = \frac{2*Precision*Recall}{Precision+Recall} = \frac{2*TP}{2*TP+FP+FN}
 \label{eq:F1}
\end{equation}

%===========================================================

\section{Experiments}
%===========================================================
%===========================================================
Three different experiments are performed to determine the effectiveness of the suggested approach and its viability for machines with fewer computation resources. Each experimental result provides a base for the subsequent experiment. These experiments are performed on a machine with an i7-8700 Intel processor, 1 TB, hard disk with 7200 RPM, 16-GB RAM (DDR4-266 MHz), without a graphics processing unit (GPU), and Windows 10 installed as OS. For the first experiment, the aim is to find the validity of the extended data set. For this experiment, the binary malware detection approach is used. All malware family feature sets are combined, generating one malware class feature set. The experimental results indicated that the extended data set can be used for multi-class classification. The second experiment is a multi-class classification experiment. This experiment aims to find the combination of the best classifier and best feature representation. A sub-aim is also to determine whether a feature fusion shall be performed. Thus, before performing feature fusion, individual feature sets are generated and fed into ten classifiers. The generated matrices contained all the malware families and benign files. It is hypothesized that each classifier will obtain a different accuracy on different feature representations. Thus, it will help to select five top classifiers to make their ensemble and also indicate the best feature sets to create their fusion. As hypothesized, the experiments resulted in different accuracies for different feature representations. However, there was no significant difference in accuracies on different feature sets. Therefore, for the next experiment, all feature sets were combined into a final matrix and fed into the top five selected algorithms. The outcome of each algorithm was used as input for a customized ensemble to determine the outcome.

%===========================================================
%===========================================================
\section{Results and Discussion}
%===========================================================
%===========================================================

%=====================================================================
\begin{figure}[!t] %for floating picture use *
	\centering
	\includegraphics[width=\linewidth,height=5cm]{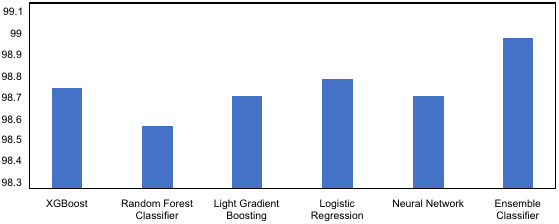}
	\caption{Classifiers' Accuracy}
	\label{fig:ClassifiersAccuracy}
\end{figure}
%=====================================================================

Malware detection and classification is typically approached as a binary classification problem. The rationale behind this idea is that many types of malware share intrinsic characteristics and boundaries, and their functionalities may be blurred or overlapped. As a result, a single file may qualify for different classes. However, the first experimental results indicate that a multi-class approach is feasible for the extended data set. The table \ref{tab:MultiClass} presents experimental results for the multi-class approach, including their training time. The best classifiers are generated with all possible parameter combinations and saved as a dictionary of accuracy with the best parameters. The dictionary is later used to get the optimal performances. The top five classifiers are selected based on training time and accuracy, with a tradeoff between the two. For example, SVM outperforms AdaBoost, \textit{k}NN, and NB on the API \textit{bi}-gram feature set, but its training time is significantly higher than other algorithms.  SVM has shown similar behavior in training time for all other feature sets. Regardless of its accuracy, SVM cannot be in the top five due to its training time. For the feature sets, the \textit{var}-gram features have provided the best results. However, all the feature representations have provided results distinct from each other with a small margin, which indicates that the feature fusion can be performed to create the input matrix. The table \ref{tab:allResults} presents the last experimental results. The results suggest that classifiers can achieve optimal performance when they use multi-feature sets. The Logistic Regression algorithm has the highest ACC at 98.819, while RF has the lowest ACC at 98.59. However, the difference between both accuracies is negligible. On the other hand, Logistics Regression has a log loss of 0.043, while RF has a log loss of 0.075, indicating that LR is a better classifier than RF.

\begin{figure} [htbp]
  \includegraphics[width=\linewidth,height=5cm]{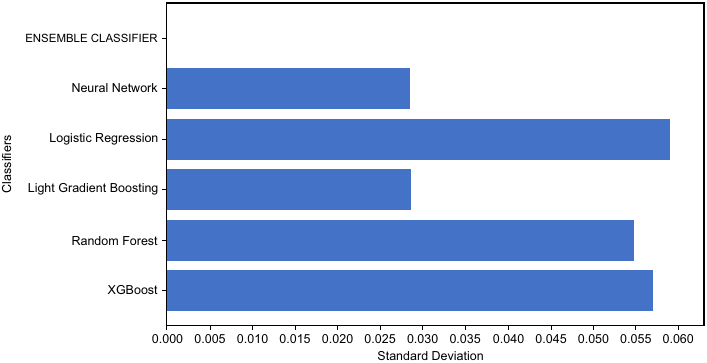}
  \caption{Classifiers' Confidence}
  \label{figMicrosoft:confidence}
\end{figure}

These top five classifiers are further combined as base algorithms to form a customized ensemble. Similar to any other algorithm, the confusion matrix for the ensemble is also generated, which is presented in Figure \ref{figSpy:ConfusionMatrixOfEnsemble}. The ensemble outperforms all its base algorithms by achieving an ACC of 99.72 and log loss of 0.01. The experimental results are presented in table \ref{tabMicrosoft:ModelComparison}. The accuracy comparison of the ensemble and its base algorithms is presented in Figure \ref{fig:ClassifiersAccuracy}. The diagram indicates that the ensemble outperforms its base algorithms. To validate the performance of the proposed ensemble algorithm, the confidence of the ensemble and its base algorithms is calculated and presented in Figure \ref{figMicrosoft:confidence}. The confidence of an algorithm is its probability of its prediction being correct. The confidence of the algorithms is calculated over 200 GB of the test data set, using the standard deviation to evaluate confidence. A lower standard deviation indicates a more robust classifier with little or no risk of overfitting. The ensemble's standard deviation is 0.00001

Although the experiments have shown optimal results, there are many challenges faced during the experiments. The primary challenge is the limited computational resources, which cannot be optimized for machine learning tasks and represent a novice user machine. Another challenge is that the data set contains encrypted files that may affect the algorithms' accuracy and confidence in finding a generalizable approach.

%===========================================================
 Moreover, to determine the validity of our approach, we have compared the suggested approach with two different approaches as follows: 
%===========================================================

\begin{table}[!htbp]
	\centering
\caption{Model Comparison}
\label{tabMicrosoft:ModelComparison}
\begin{tabular}{|l|c|c|}
\hline 
         & \textbf{Winner's Model} & \textbf{Proposed Model} \\
         \hline 
         
Accuracy & 99.50\%        & 99.72\%        \\ \hline
Log loss & 0.002          & 0.01          \\
\hline
\end{tabular}
\end{table}

%=====================================================================

\paragraph{Comparison with the Competition's Winner}
%\subsection{Comparison with state-of-the-art}
The private and public ladder boards of the competition are consulted to find the competition's winner. According to the private ladder board, the competition winner team, i.e., "\textit{say NOOOOO to overfittttting}"\footnote{https://github.com/xiaozhouwang/kaggle\_Microsoft\_Malware} has reported 0.0023 and 0.0028 multiclass log loss. The winning team's approach can be divided into two distinct parts, i.e., feature engineering and modeling. The usage of the XGBoost algorithm influences both parts. For feature engineering, three types of features are extracted, i.e., OpCode \textit{n}-gram and their count, segment line count, and ASM file pixel intensity features. From the OpCode \textit{n}-gram, they also generated \textit{bi}-gram, tri-gram, and \textit{quad}-gram. However, OpCode count based on frequency, segment count, and ASM file pixel intensity are novel features and provide the best results. For the feature selection, they used information gain, random forest, and XGBoost at different stages. For the modeling, authors have used different algorithms and techniques, i.e., Random Forest, Naive Bayes, Neural Network, Gradient Boosting, and semi-supervised learning. They also used cross-validation to select their model. For the computation resources, the winning team used Google Compute Engine (instance with 104G memory with 16 CPUs) for the hardware. 

%==========================================================
%\begin{figure} [!t]
 % \includegraphics[width=\linewidth]{confidence.pdf}
 % \caption{Classifiers' Confidence}
 % \label{figMicrosoft:confidence}
%\end{figure}
%==========================================================

Unfortunately, due to the competition's rules, only log loss is reported by the participants. Thus, it is hard to have a thorough comparison of the approaches. However, the approach can be compared for the data set used,  features used, modeling, log loss, and computation resources. The main difference in the overall approach is that the winning team used encrypted and unencrypted samples, and the third feature, i.e., ASM pixel intensity, is used because of encrypted samples. However, they have not used benign examples, which may differ in structure.
Thus, they have used the given nine classes for their data set. During the experiment, they used many hard coded hyper-parameters relevant to these nine classes. In comparison, the suggested approach has used ten different classes, and no hard coded hyper-parameters are used for experimentation. Moreover, the winning team has used features that may only be valuable for the competition.  Thus, the generalization of the approach is questionable. The file size may affect the approach and change the OpCode or instruction count and other features. The other problem is that their approach may not help track a characteristic of the relevant file. Another issue is that their approach only provides structural information, which may not help for further analysis. In comparison, the suggested approach uses four features, i.e., \textit{quad}-gram, \textit{var}-gram, API, and DLL. These features can be tracked back to the original files in the data set to understand the functionality of the file and generalize the approach. Moreover, the \textit{variable-length}-grams can provide significant information about the functionality of the file. Other features, i.e., DLL and API, can also be traced back to the original files for further analysis. Moreover, the winning team has performed incremental feature selection at an increment of 1000. However, the proposed approach combines common feature selection measures with a customized Backward feature selection using a threshold.  
For the modeling part of the approach, the ensemble approach is suggested, and it has shown better results than the winning team's reported results. Moreover, the winning team has relied both on supervised and semi-supervised learning algorithms. However, the proposed approach has relied on supervised learning algorithms. For the log loss, the winning team has shown a lower log loss as presented in table \ref{tabMicrosoft:ModelComparison}.  However, the winning team has Performed thousands of iterations to reduce the loss of the classifiers, while the proposed method performs moderate iterations due to resource constraints. For the last part, i.e., computational resources, the proposed approach is suitable for running and updating with minimum computational resources. In conclusion, the proposed approach can be generalized, while the winning team's approach is competition-focused and cannot be generalized.
%==========================================================
%==========================================================
\paragraph{Comparison with Dataset's Owners}
The proposed model is compared against another state-of-the-art study, which has similar characteristics \citep{ahmadi2016novel}. %The selected study may be considered as a first study on the data set, and 
It is worth noting that one of the authors is part of the team that released the data set. The study can be compared against the used features, classification models, and computational complexity. For the features, the authors have used both hexadecimal and ASM based features. For the hex-dump, they have used \textit{uni}-gram, which may not give any valuable information for the analysis. They are also hard to track for further analysis. They also used meta-data and string length information. These features may be helpful for a particular data set and may not be used for generalizing the approach. From the disassembled files, the authors have extracted meta-data, frequencies of different features, and the ratio of instructions as features. These features also represent a similar problem, i.e., they may vary with changes in the data set and can not generalize the approach.
In contrast to these features, the proposed features in our approach may be tracked for further analysis and may not vary with the data set, especially \textit{variable-length} \textit{n}-grams. 
Moreover, these features capture a function; thus, they may also indicate the function's objective. Moreover, the features used in the selected study may take more time to extract and require a lot of read/write operations. In contrast, the proposed method extracts the basic features once, and later extracted features are used to derive the required features for input. This also helps upgrade the input matrix in less time if a model needs to be retrained with more examples.  

For the feature selection, 
authors have used a modified forward step-wise selection, which starts with zero features and gradually augments features set with features with minimum log loss from a particular category. This study performs feature selection step by step, i.e., primary and secondary feature selection. Later, in the secondary feature selection, this study also exploits the potential of all feature selection strategies and use methods, which are often favored in literature. This increased the feature selection time; however, it ensures that only valuable features are part of the input. 

For the classification, both methods rely on an ensemble; the article used XGBoost with bagging, and this study uses an ensemble of different classifiers. This study also shows that XGBoost can be used without generating an ensemble with the suggested features and achieve good results. 
\section{Conclusion}
Automatic malware detection and classification to respective malware families may help human experts save time and launch timely responses. This paper presents an automated malware detection method that classifies the detected malware to the respective malware family. For classification, static features are extracted from disassembled binary files. A customized feature selection method is suggested to reduce feature dimensionality before the feature fusion. The reduced data set is fed into different algorithms, including a customized ensemble algorithm based on algorithm fusion. The experimental results suggest that feature fusion and algorithm fusion may help in malware detection and family classification. In the future, we plan to investigate a feature fusion between the features from the hexadecimal data set and the disassembled data set. We also plan to increase the number of families and include encrypted samples.
\section*{Declarations}
The authors did not receive support from any organization for the submitted work. Moreover, the authors have no relevant financial or non-financial interests to disclose.
\bibliography{microsoft}

@inproceedings{Noelia2007, 
	author={S{\'a}nchez-Maro{\~{n}}o, Noelia and Alonso-Betanzos, Amparo and Tombilla-Sanrom{\'a}n, Mar{\'i}a},
	editor={Yin, Hujun and Tino, Peter and Corchado, Emilio and Byrne, Will and Yao, Xin},
	title={Filter Methods for Feature Selection -- A Comparative Study},
	booktitle={Intelligent Data Engineering and Automated Learning - IDEAL 2007},
	series={Lecture Notes in Computer Science},
	vol={4881}, 
	year={2007},
	publisher={Springer Berlin Heidelberg},
	pages={178--187},
	isbn={978-3-540-77226-2}
}

@book{Malware_Analysis_Book_2012,
	author = {Sikorski, Michael and Honig, Andrew},
	title = {Practical Malware Analysis: The Hands-On Guide to Dissecting Malicious Software},
	year = {2012},
	isbn = {1593272901},
	publisher = {No Starch Press},
	address = {USA},
	edition = {1st},
}

@inproceedings{Noelia_2009,
	author={S{\'a}nchez-Maro{\~{n}}o, Noelia and Alonso-Betanzos, Amparo and Calvo-Est{\'e}vez, Rosa M.},
	editor={Cabestany, Joan and Sandoval, Francisco and Prieto, Alberto and Corchado, Juan M.},
	title={A Wrapper Method for Feature Selection in Multiple Classes Datasets},
	booktitle={Bio-Inspired Systems: Computational and Ambient Intelligence},
	year={2009},
	publisher={Springer Berlin Heidelberg},
	pages={456--463}
}

@inproceedings{shahzad_adware_2011,
	title = {Accurate Adware Detection Using Opcode Sequence Extraction},
	isbn = {978-1-4577-0979-1},
	booktitle = {Proceedings of the Sixth International Conference on Availability, Reliability and Security},
	publisher = {{IEEE}},
	author = {Shahzad, Raja Khurram and Lavesson, Niklas and Johnson, Henric},
	year = {2011},
	pages = {189--195}
}

@book{kuhn2013applied,
  title={Applied Predictive Modeling},
  author={Kuhn, M. and Johnson, K.},
  isbn={9781461468493},
  series={SpringerLink : B{\"u}cher},
  year={2013},
  publisher={Springer New York}
}

@inproceedings{ahmadi2016novel,
	title={Novel Feature Extraction, Selection and Fusion for Effective Malware Family Classification}, 
	author={Mansour Ahmadi and Dmitry Ulyanov and Stanislav Semenov and Mikhail Trofimov and Giorgio Giacinto},
	year={2016},
	booktitle={Proceedings of the Sixth ACM Conference on Data and Application Security and Privacy (CODASPY '16)},
	publisher={ACM},
	pages={183-194}, 
}

@Inbook{Vovk2015,
	author={Vovk, Vladimir},
	editor={Beklemishev, Lev D.and Blass, Andreas and Dershowitz, Nachum and Finkbeiner, Bernd and Schulte, Wolfram},
	title={The Fundamental Nature of the Log Loss Function},
	bookTitle={Fields of Logic and Computation II: Essays Dedicated to Yuri Gurevich on the Occasion of His 75th Birthday},
	year={2015},
	publisher={Springer International Publishing},
	pages={307--318},
	isbn={978-3-319-23534-9},
	doi={10.1007/978-3-319-23534-9_20}
}

@inproceedings{Shahzad_consensus_2015,
	author={Shahzad, Raja Khurram and Fatima, Mehwish and Lavesson, Niklas and Boldt, Martin},
	editor={Pardalos, Panos and Pavone, Mario and Farinella, Giovanni Maria and Cutello, Vincenzo},
	title={Consensus Decision Making in Random Forests},
	booktitle={Machine Learning, Optimization, and Big Data},
	year={2015},
	publisher={Springer International Publishing},
	pages={347--358},
	isbn={978-3-319-27926-8}
}

@inproceedings{Chen_2016,
	author = {Chen, Tianqi and Guestrin, Carlos},
	title = {XGBoost: A Scalable Tree Boosting System},
	year = {2016},
	isbn = {9781450342322},
	publisher = {Association for Computing Machinery},
	booktitle = {Proceedings of the 22nd ACM SIGKDD International Conference on Knowledge Discovery and Data Mining},
	pages = {785–794},
	numpages = {10},
	series = {KDD '16}
}

@book{Witten2016,
	author = {Witten, Ian H. and Frank, Eibe and Hall, Mark A. and Pal, Christopher J.},
	title = {Data Mining, Fourth Edition: Practical Machine Learning Tools and Techniques},
	year = {2016},
	isbn = {0128042915},
	publisher = {Morgan Kaufmann Publishers Inc.},
	address = {San Francisco, CA, USA},
	edition = {4th}
}

@inproceedings{Chen_2017,
	author={Chen, Lingwei and Ye, Yanfang and Bourlai, Thirimachos},
	booktitle={European Intelligence and Security Informatics Conference (EISIC)}, 
	title={Adversarial Machine Learning in Malware Detection: Arms Race between Evasion Attack and Defense}, 
	year={2017},
	pages={99-106},
	doi={10.1109/EISIC.2017.21}
}

@article{ronen2018microsoft,
	author = {Royi Ronen and Marian Radu and Corina Feuerstein and Elad Yom{-}Tov and Mansour Ahmadi},
	title        = {Microsoft Malware Classification Challenge},
	journal      = {CoRR: a computing research repository},
	volume       = {abs/1802.10135},
	year         = {2018}
	
}

@inproceedings{Hu_2019,
	author={Hu, Yen-Hung Frank and Ali, Abdinur and Hsieh, Chung-Chu George and Williams, Aurelia},
	booktitle={2019 SoutheastCon}, 
	title={Machine Learning Techniques for Classifying Malicious API Calls and N-Grams in Kaggle Data-set}, 
	year={2019},
	pages={1-8},
	doi={10.1109/SoutheastCon42311.2019.9020353}
}

@inproceedings{Unal_20,
	author = {Unal, Ugur and Yenido\u{g}an, I\c{s}{\i}l and Dag, Hasan and Cayir, Aykut},
	year = {2019},
	booktitle = {2019 International Conference on Data Science, Machine Learning and Statistics (DMS-2019)}, 
	month = {09},
	title = {Use Case Study: Data Science Application for Microsoft Malware Prediction Competition on Kaggle}
}

@inproceedings{Gibert_Orthurs_20,
	author={Gibert, Daniel and Mateu, Carles and Planes, Jordi},
	booktitle={2020 International Joint Conference on Neural Networks (IJCNN)}, 
	title={Orthrus: A Bimodal Learning Architecture for Malware Classification}, 
	year={2020},
	pages={1-8},
	doi={10.1109/IJCNN48605.2020.9206671}
}

@inproceedings{Ding_Self_20,
	title={Malware Classification on Imbalanced Data through Self-Attention},
	author={Yu Ding and Shupeng Wang and Jian Xing and Xiaoyu Zhang and ZiSen Oi and Ge Fu and Qian Qiang and Haoliang Sun and Jianyu Zhang},
	booktitle={2020 IEEE 19th International Conference on Trust, Security and Privacy in Computing and Communications (TrustCom)},
	year={2020},
	pages={154-161}
}

@inproceedings{Chen_20,
	author={Chen, Chin-Wei and Su, Ching-Hung and Lee, Kun-Wei and Bair, Ping-Hao},
	booktitle={2020 22nd International Conference on Advanced Communication Technology (ICACT)}, 
	title={Malware Family Classification using Active Learning by Learning}, 
	year={2020},
	pages={590-595},
	doi={10.23919/ICACT48636.2020.9061419}
}

@inproceedings{Mark_21,
	author={Mark Sokolov and Nic Herndon},
	title={Predicting Malware Attacks using Machine Learning and AutoAI},
	booktitle={Proceedings of the 10th International Conference on Pattern Recognition Applications and Methods - Volume 1: ICPRAM,},
	year={2021},
	pages={295-301},
	publisher={SciTePress},
	organization={INSTICC},
	doi={10.5220/0010264902950301},
	isbn={978-989-758-486-2},
}

@Article{Attaluri2009,
	author={Attaluri, Srilatha 	and McGhee, Scott and Stamp, Mark},
	title={Profile hidden Markov models and metamorphic virus detection},
	journal={Journal in Computer Virology},
	year={2009},
	month={May},
	day={01},
	volume={5},
	number={2},
	pages={151-169},
	issn={1772-9904},
	doi={10.1007/s11416-008-0105-1}
}

@article{shabtai_2009,
	title = {Detection of Malicious Code by applying Machine Learning Classifiers on Static Features: A {State-of-the-Art} Survey},
	volume = {14},
	issn = {1363-4127},
	shorttitle = {Detection of Malicious Code by applying Machine Learning Classifiers on Static Features},
	number = {1},
	journal = {Information Security Technical Report},
	author = {Shabtai, Asaf and Moskovitch, Robert and Elovici, Yuval and Glezer, Chanan},
	year = {2009},
	pages = {16--29}
}

@article{Rie2014,
	Title = {Learning Nonlinear Functions Using Regularized Greedy Forest},
	Author = {Johnson, Rie and Tong Zhang},
	Number = {5},
	Volume = {36},
	Month = {May},
	Year = {2014},
	Journal = {IEEE transactions on pattern analysis and machine intelligence},
	ISSN = {0162-8828},
	Pages = {942—954},
	doi = {10.1109/TPAMI.2013.159}
}

@article{Liu_2017,
	author={Liu, Liu and Wang, Bao-sheng and Yu, Bo and Zhong, Qiu-xi},
	title={Automatic malware classification and new malware detection using machine learning},
	journal={Frontiers of Information Technology {\&} Electronic Engineering},
	year={2017},
	month={Sep},
	day={01},
	volume={18},
	number={9},
	pages={1336-1347},
	issn={2095-9230},
	doi={10.1631/FITEE.1601325}
}

@article{Sagi_Ensemble_learning,
	author = {Sagi, Omer and Rokach, Lior},
	title = {Ensemble learning: A survey},
	journal = {WIREs Data Mining and Knowledge Discovery},
	volume = {8},
	number = {4},
	pages = {e1249},
	year = {2018}
}

@article{Dong_ensemble_2020,
	author={Dong, Xibin	and Yu, Zhiwen	and Cao, Wenming	and Shi, Yifan	and Ma, Qianli},
	title={A survey on ensemble learning},
	journal={Frontiers of Computer Science},
	year={2020},
	volume={14},
	number={2},
	pages={241-258},
	issn={2095-2236}
}

@article{chowdhury_variable_2020,
	title = {Variable selection strategies and its importance in clinical prediction modelling},
	volume = {8},
	issn = {2305-6983},
	doi = {10.1136/fmch-2019-000262},
	number = {1},
	journal = {Family medicine and community health},
	author = {Chowdhury, Mohammad Ziaul Islam and Turin, Tanvir C},
	month = {feb},
	year = {2020},
	note = {Publisher: BMJ Publishing Group}
	
}

@article{gibert_rise_of_machine_learning_2020,
	title = {The rise of machine learning for detection and classification of malware: Research developments, trends and challenges},
	journal = {Journal of Network and Computer Applications},
	volume = {153},
	pages = {102526},
	year = {2020},
	issn = {1084-8045},
	author = {Daniel Gibert and Carles Mateu and Jordi Planes}
}

@article{Euh_20,
	author={Euh, Seoungyul and Lee, Hyunjong and Kim, Donghoon and Hwang, Doosung},
	journal={IEEE Access}, 
	title={Comparative Analysis of Low-Dimensional Features and Tree-Based Ensembles for Malware Detection Systems}, 
	year={2020},
	volume={8},
	pages={76796-76808},
	doi={10.1109/ACCESS.2020.2986014}
}

@inproceedings{Pan_20,
	author = {{Pan}, Qiangjian and {Tang}, Weiliang and {Yao}, Siyue},
	title = "{The Application of LightGBM in Microsoft Malware Detection}",
	booktitle = {Journal of Physics Conference Series},
	year = {2020},
	series = {Journal of Physics Conference Series},
	volume = {1684},
	month = {nov},
	pages = {012041},
	doi = {10.1088/1742-6596/1684/1/012041}
	
}

@article{Sahin2021,
	author={{\c{S}}ahin, Durmu{\c{s}} {\"O}zkan and Kural, O{\u{g}}uz Emre and Akleylek, Sedat and K{\i}l{\i}{\c{c}}, Erdal},
	title={A novel Android malware detection system: adaption of filter-based feature selection methods},
	journal={Journal of Ambient Intelligence and Humanized Computing},
	year={2021},
	month={Jul},
	issn={1868-5145},
	doi={10.1007/s12652-021-03376-6}
}

@article{WANG_2022,
	title = {Malicious code classification based on opcode sequences and textCNN network},
	author = {Qianhui Wang and Quan Qian},
	journal = {Journal of Information Security and Applications},
	volume = {67},
	pages = {103151},
	year = {2022},
	issn = {2214-2126},
	doi = {https://doi.org/10.1016/j.jisa.2022.103151}
}

@article{ZHU_2022,
	title = {N-gram MalGAN: Evading machine learning detection via feature n-gram},
	journal = {Digital Communications and Networks},
	volume = {8},
	number = {4},
	pages = {485-491},
	year = {2022},
	issn = {2352-8648},
	doi = {https://doi.org/10.1016/j.dcan.2021.11.007},
	author = {Enmin Zhu and Jianjie Zhang and Jijie Yan and Kongyang Chen and Chongzhi Gao}
}

@article{LI_2022,
	title = {Malware classification based on double byte feature encoding},
	journal = {Alexandria Engineering Journal},
	volume = {61},
	number = {1},
	pages = {91-99},
	year = {2022},
	issn = {1110-0168},
	doi = {https://doi.org/10.1016/j.aej.2021.04.076},
	url = {https://www.sciencedirect.com/science/article/pii/S1110016821003185},
	author = {Lin Li and Ying Ding and Bo Li and Mengqing Qiao and Biao Ye}
}

@article{GIBERT_2022,
	title = {Fusing feature engineering and deep learning: A case study for malware classification},
	journal = {Expert Systems with Applications},
	volume = {207},
	pages = {117957},
	year = {2022},
	issn = {0957-4174},
	doi = {https://doi.org/10.1016/j.eswa.2022.117957},
	author = {Daniel Gibert and Jordi Planes and Carles Mateu and Quan Le}
}

@article{GUO_2023,
	title = {A review of Machine Learning-based zero-day attack detection: Challenges and future directions},
	journal = {Computer Communications},
	volume = {198},
	pages = {175-185},
	year = {2023},
	issn = {0140-3664},
	doi = {https://doi.org/10.1016/j.comcom.2022.11.001},
	url = {https://www.sciencedirect.com/science/article/pii/S0140366422004248},
	author = {Yang Guo}
}

@article{TANG_2023,
	title = {BHMDC: A byte and hex n-gram based malware detection and classification method},
	journal={Computers {\&} Security},
	volume = {128},
	pages = {103118},
	year = {2023},
	issn = {0167-4048},
	doi = {https://doi.org/10.1016/j.cose.2023.103118},
	author = {Yonghe Tang and Xuyan Qi and Jing Jing and Chunling Liu and Weiyu Dong}
	
}

@article{NAEEM_2023,
	title = {Development of a deep stacked ensemble with process based volatile memory forensics for platform independent  malware detection and classification},
	journal = {Expert Systems with Applications},
	volume = {223},
	pages = {119952},
	year = {2023},
	issn = {0957-4174},
	doi = {https://doi.org/10.1016/j.eswa.2023.119952},
	author = {Hamad Naeem and Shi Dong and Olorunjube James Falana and Farhan Ullah}
}

@article{ALANI_2023,
	title = {XMal: A lightweight memory-based explainable obfuscated-malware detector},
	journal = {Computers {\&} Security},
	volume = {133},
	pages = {103409},
	year = {2023},
	issn = {0167-4048},
	doi = {https://doi.org/10.1016/j.cose.2023.103409},
	author = {Mohammed M. Alani and Atefeh Mashatan and Ali Miri}
}

@article{RUSTAM_2023,
	title = {Malware detection using image representation of malware data and transfer learning},
	journal = {Journal of Parallel and Distributed Computing},
	volume = {172},
	pages = {32-50},
	year = {2023},
	issn = {0743-7315},
	doi = {https://doi.org/10.1016/j.jpdc.2022.10.001},
	author = {Furqan Rustam and Imran Ashraf and Anca Delia Jurcut and Ali Kashif Bashir and Yousaf Bin Zikria}
}

@article{JOYCE_2023,
	title = {MOTIF: A Malware Reference Dataset with Ground Truth Family Labels},
	journal = {Computers  {\&} Security},
	volume = {124},
	pages = {102921},
	year = {2023},
	issn = {0167-4048},
	doi = {https://doi.org/10.1016/j.cose.2022.102921},
	author = {Robert J. Joyce and Dev Amlani and Charles Nicholas and Edward Raff},
}

@article{MIMURA_2023,
	title = {Impact of benign sample size on binary classification accuracy},
	journal = {Expert Systems with Applications},
	volume = {211},
	pages = {118630},
	year = {2023},
	issn = {0957-4174},
	doi = {https://doi.org/10.1016/j.eswa.2022.118630},
	author = {Mamoru Mimura}
}
\end{document}